
%
\magnification=\magstep1
\vsize=23truecm
\hsize=15.5truecm
\hoffset=.2truecm
\voffset=.8truecm
\parskip=.2truecm

\font\ti=cmbx10 scaled\magstep1
\font\eightrm=cmr8
\font\ninerm=cmr9
\def\br{\hfill\break\noindent}

\def \ot {\otimes}
\def \g5{\gamma_5}
\def \ra{\rightarrow}
\def \l4{\Bigl( {\rm Tr}( KK^*)^2-({\rm Tr}KK^*)^2\Bigr)}
\def \bp{\oplus }
\def \k2{{\rm Tr}KK^*}
\def \slash#1{/\kern -6pt#1}
\def \di{\slash{\partial}}
\def \ol {\overline }
\pageno=0
\def \bt {\otimes}
%
%
\baselineskip=.5truecm
\footline={\hfill}
{\hfill ZU-TH- 30/1992}
\vskip.1truecm
{\hfill ETH/TH/92-41 }
\vskip.2truecm
{\hfill 3 September 1992}
\vskip2.1truecm
\centerline{\ti Grand Unification in Non-Commutative Geometry }
\vskip1.2truecm
\centerline{  A. H. Chamseddine$^{1}$ \footnote*
{\ninerm Supported by the Swiss National Foundation (SNF)}
G. Felder$^{2}$ and J. Fr\"ohlich$^{3}$ }
\vskip.8truecm
\centerline{$^{1}$ Theoretische Physik, Universit\"at Z\"urich, CH
8001 Z\"urich Switzerland}
\centerline{$^{2}$  Department of Mathematics, ETH,
CH 8092 Z\"urich Switzerland}
\centerline{$^{3}$ Theoretische Physik, ETH, CH 8093 Z\"urich Switzerland}
\vskip1.2truecm

\centerline{\bf Abstract}
\vskip.5truecm

\noindent
The formalism of non-commutative geometry of A. Connes is used to
construct models in particle physics. The physical space-time is
taken to be a product of a continuous four-manifold by a discrete
set of points. The treatment of Connes is modified in such a way
that the basic algebra  is defined over the space of matrices,
and the breaking mechanism is planted in the Dirac operator.
This mechanism is then applied to three examples.
In the first example the discrete space consists of two points,
and the two algebras are taken respectively to be those of $2\times 2$ and
$1\times 1$ matrices. With the Dirac operator containing the
vacuum breaking $SU(2)\times U(1)$ to $U(1)$, the model is shown to
correspond to the standard model. In the second example the discrete
space has three points, two of the algebras are identical and consist
of $5\times 5$ complex matrices, and the third algebra consists
of functions. With an appropriate Dirac operator this model is
almost identical to the minimal $SU(5)$ model of Georgi and Glashow.
The third and final example is the left-right symmetric model
$SU(2)_L\times SU(2)_R\times U(1)_{B-L}.$
\vfill

\eject
\baselineskip=.6truecm
\footline={\hss\eightrm\folio\hss}

\centerline{}
\vskip1.1truecm
{\bf\noindent 1. Introduction}
\vskip.2truecm

\noindent
At present energies the standard model of electroweak
interactions has passed all experimental tests. One of the essential
ingredients of this model is the Higgs field. The presence of
the Higgs field is required to break the gauge symmetry spontaneously.
{}From the four-dimensional point of view, there is no apparent geometrical
reason for the Higgs field. Although there are some possible candidates,
e.g. a Kaluza-Klein theory or a compactified string model, there
are no compelling models yet. A new picture was put forward
recently by Connes [1-2], where the experimental validity of the standard
model was taken as an indication for a non-commutative picture
of space-time. Space-time is taken to be a product of a continuous
Euclidean manifold $M_4$ by a discrete "two-point" space. The fibers
in the two copies of space are taken to be $U(1)$ and $SU(2)$ respectively.
The vector potential defined in this space will have as
components $U(1)$ and $SU(2)$ gauge fields along the continuous directions
and the scalar Higgs field along the discrete directions.
Therefore, non-commutative geometry offers a geometrical picture
for the unification of the gauge and Higgs field. The advantage of
this approach over the Kaluza-Klein approach is that there is no
truncation of any physical modes, while, in the latter, an infinite
number of massive modes is truncated. In this formalism it was
shown by Connes and Lott [3-4], and elaborated upon in great detail
by Kastler [5],
on how to construct the standard model. Inclusion of the $SU(3)$
strong interaction  proved to be more difficult and was only
achieved  recently [3-5]. Other constructions were proposed by different
authors, such as Coquereux et al , Dubois-Violette et al.  and Balakrishna
et al. [6], but
they lack a compelling geometrical structure and will not be followed here.

It is usually expected  that the standard model [7] will
be replaced by a different theory at higher energies
which one hopes to be more unified. In particular, the grand unified
theories (GUTs) seem to provide (in their supersymmetric forms)
acceptable models. The problem that will be addressed in
this paper is to find a way to build GUTs models and other
possible models at energies higher than the weak scale, within
the non-commutative picture.

The strategy adopted in references [1-5] is only appropriate in the
case of a product symmetry such as $SU(2)\times U(1)$. If one
follows this strategy without modification,
many difficulties will be encountered and no
phenomonologically successful grand-unified model can
be built. This strategy also excludes a single gauge group. To
explore other  possibilities, we note that a typical GUT
involves at least two scales: the grand unification scale, where
the three coupling constants of $SU(3)$, $SU(2)$ and $U(1)$
coincide, and the electroweak scale. In addition, there could be
intermediate scales. By choosing space-time to be a product of a
continuous four-dimensional Riemannian manifold by a discrete
set of points we immediately see that the simplest possibility
for the choice of a Dirac operator including  more than one
scale is to take the discrete space to consist of three points.
This is the situation we shall be mostly interested in, although the
extension to a discrete space of $N$ points is straightforward.
By generalizing
the algebra of functions to be given by a direct sum of algebras of
matrix-valued functions and by planting the symmetry
breaking mechanism in the Dirac
operator, it will turn out to be possible to construct unified models.

The plan of this paper is as follows: In section 2, we modify
the prescription of Connes [1-2] in such a way that the discrete space
consists of three points generalizable to $N$ points. We introduce
the idea of planting the symmetry breaking  in the Dirac operator
and prove that this does not break gauge invariance. In section 3,
and as a warm up, we apply this prescription to construct the
standard model of the electorweak interactions.
In section 4, we construct the $SU(5)$ model and obtain the minimal
model (apart from an extra Higgs singlet) of Georgi an Glashow [8].
 In section 5, we
construct the model $SU(2)_L\times SU(2)_R\times U(1)_{B-L}$ of
Pati and Mohapatra [9] by taking the discrete space to consist
of four points.
Section 6, contains our conclusions.


{\bf \noindent 2. A new prescription for model building  }
\vskip .2truecm
\noindent
Consider a model of non-commutative geometry consisting
of the  triple $({\cal A},h,D)$,
where $h$ is a Hilbert space, ${\cal A}$ is an involutive algebra
of operators on $h$, and $D$ is an unbounded self-adjoint
operator on $h$. An example important for our construction is
the following one:
Let $X$ be a compact Riemannian spin-manifold,
${\cal A}_1$ the algebra of functions on $X$, and
$(h_1,D_1,\Gamma_1 )$ the Dirac-$K$ cycle  with $h_1\equiv L^2 (X,
\sqrt g d^d x )$ on ${\cal A}_1$. Let
$({\cal A}_2,h_2,D_2)$ be given by ${\cal A}_2=M_n(C)\bp
M_p(C)\bp M_q(C) $, where $M_n(C)$ is the set of all
$n\times n $ matrices and $h_2=h_{2,1}\oplus h_{2,2} \oplus
h_{2,3} $ where $h_{2,1}$  $h_{2,2}$ and $h_{2,3}$ are the Hilbert
spaces $C^n$, $C^p $ and $C^q$, respectively.
Then ${\cal A}$ and $D$ are taken to be
$$\eqalignno{
{\cal A}&={\cal A}_1\otimes {\cal A}_2 \cr
D&=D_1\otimes 1 +\Gamma_1 \otimes D_2 &(2.1)\cr}
$$

To every $f\epsilon {\cal A}$ we associate a triplet $(f_1, f_2,
f_3)$ of matrix-valued functions on $X$, where $f_1$, $f_2$, and $f_3$ are
 $n\times n$, $p\times p$, and $q\times q$ matrices, respectively. The
decomposition of $h_2$ corresponds to the decomposition
$h=h_1\oplus h_2 \oplus h_3$ for which the action of $f$ is
block-diagonal
$$
f\rightarrow {\rm diag }(f_1,f_2,f_3). \eqno(2.2)
$$
In this decomposition, the operator $D$ becomes
$$
 D=\bordermatrix {&n&p&q\cr
n&\di\bt 1&\g5 \bt M_{12}&\g5 \bt M_{13} \cr
p&\g5 \bt M_{21} &\di \bt 1&\g5 \bt M_{23}\cr
q&\g5 \bt M_{31} &\g5 \bt M_{32} & \di \bt 1 \cr}\eqno(2.3)
$$
where $M_{mn}^* =M_{nm}$ and $m,n=1,2,3, m\not=n$ .
The gamma matrices we use satisfy: $\gamma_a^*=-\gamma_a $,
$\{\gamma_a ,\gamma_b \}=-2\delta_{ab}$,$\gamma_5=\gamma_1\gamma_2
\gamma_3\gamma_4 $, $\gamma_5^*=\gamma_5 $, and
$g_{ab}=-\delta_{ab} $ is the Euclidean metric.

An important difference between our approach and the prescription
given by Connes et al. is that they choose all the matrices $M_{mn}$
to be of the same size, i.e. $n=p=q$, and proportional to the
identity matrix. In our approach they can be general matrices and do
$\underline { \rm not}$  commute with elements of ${\cal A} $.
The novel idea that we
will advance is that the matrices $M_{mn}$ of the model
determine the tree level vacuum-expectation values of Higgs fields
 and the desired symmetry
breaking scheme. This
modification  allows us, first, to simplify the construction of
the standard model and then go beyond this model to grand unification
models.

Let $E$ be a vector bundle characterized by the vector space
$\cal E$ of its sections.  We shall consider the example where
 ${\cal E}={\cal A}$.
 Let $\rho $ be a self-adjoint element
in the space, $\Omega^1 ({\cal A})$, of one forms
$$
\rho =\sum_{i} a^i db^i, \qquad (d1=0)  \eqno(2.4)
$$
where $\Omega^* ({\cal A})=\oplus_{n=0}^{\infty} \Omega^n ({\cal
A})$ is the universal differential algebra, with $\Omega^0
({\cal A})={\cal A}$; See [1]. (The space $\Omega^n ({\cal A})$
plays the role of $n$-forms in non-commutative geometry.)

An involutive representation of $\Omega^* ({\cal A}) $ is provided by the map
$\pi : \Omega^* ({\cal A})\rightarrow B(h) $  defined by
$$
\pi (a_0da_1...da_n)=a_0[D,a_1][D,a_2]...[D,a_n]  \eqno(2.5)
$$
where $B(h)$ is the algebra of bounded operators
on $h$.
 The image of the one-form $\rho $ is
$$
\pi (\rho ) =\sum_i a^i[D, b^i],  \eqno(2.6)
$$
where the elements $a^i$ and $b^i$ are represented by
$$\eqalign {
a^i &\rightarrow {\rm diag } (a_1^i, a_2^i, a_3^i)\cr
b^i &\rightarrow {\rm diag } (b_1^i, b_2^i, b_3^i) \cr}\eqno(2.7)
$$
interpreted as bounded operators on $h$.
The product $a^i[D,b^i] $ is  defined in terms of standard multiplication.
Using the expression of eq.(2.3) for $D$, the commutator $[D,
b]$ can be easily evaluated and is given by
$$
[D, b]=\pmatrix {\di b_1 &\g5 \bt (M_{12}b_2-b_1M_{12}) &\g5 \bt (M_{13}
b_3-b_1M_{13})\cr
\g5 \bt (M_{21}b_1-b_2 M_{21}) &\di b_2 & \g5 \bt (M_{23}b_3-b_2M_{23})\cr
\g5 \bt (M_{31}b_1-b_3M_{31}) & \g5 \bt (M_{32}b_2-b_3M_{32})
&\di b_3 \cr} \eqno(2.8)
$$
Inserting eq.(2.8) in eq.(2.6), we obtain
$$
\pi (\rho )=\pmatrix{A_1 &\g5 \bt \phi_{12} &\g5 \bt \phi_{13} \cr
\g5 \bt \phi_{21} & A_2 &\g5 \bt \phi_{23} \cr
\g5 \bt \phi_{31} & \g5 \bt \phi_{32} &A_3 \cr} \eqno(2.9).
$$
where the new variables $A$ and $\phi $ are functions of the
$a's$ and the $b's$ given by
$$\eqalign{
A_m &=\sum_i a_m^i\di b_m^i, \qquad m=1,2,3, \cr
\phi_{mn}&=\sum_i a_m^i (M_{mn}b_n^i -b_m^iM_{mn}),\qquad
m\not= n, \cr}\eqno(2.10)
$$
and satisfy $A_m^{\ast }=A_m$ and $\phi_{mn}^{\ast }=\phi_{nm} $.

The two-form $d\rho $ is:
$$
d\rho =\sum_i da^i db^i  \eqno(2.11)
$$
and its immage under the involutive representation $\pi $ is
given by
$$
\pi (d\rho )=\sum_i [D, a^i][D, b^i]  \eqno(2.12)
$$
At this point we can address the question of gauge invariance.
If one wishes for the action of a spinor field
$$
 <\Psi ,(D+\pi (\rho ))\Psi > \eqno(2.13)
$$
to be invariant under the transformation $\Psi \rightarrow
\ ^g\Psi =g\Psi $, where $g\epsilon {\cal A} $  satisfies
$g\epsilon U({\cal A})=\{ g\epsilon {\cal A} \vert g^*g=1 \}$
is unitary,
 then $\rho $
must transform inhomogeneously according to
$$
\ ^g\rho =g\rho g^* +gdg^*  \eqno(2.14)
$$
This is consistent with the definition of $\rho $ in eq.(2.4):
$$
\ ^g\rho =\sum_i (ga^i)d(b^ig^*)-g\bigl( (\sum_i a^ib^i)-1 \bigr)dg^*
\eqno(2.15)
$$
where the second term could be included in the first term by
enlarging the set of the $a^i$'s and $b^i$'s. It is possible to define
gauge transformations explicitly
on the elements $a^i$ and $b^i$ :
$$\eqalign{
a^i&\rightarrow \ ^ga^i=ga^i \cr
b^i&\ra \ ^gb^i=b^i g^* } \eqno(2.16)
$$
provided one imposes the constraint
$$
\sum_i a^i b^i =1     \eqno(2.17)
$$
This is no loss in generality, as the field
$\sum_i a^i b^i  $ is  independent. We shall use the constraint
(2.17) and the transformations (2.16) when convenient.
Similarly, the transformation of $d\rho $ could be
easily derived to be
$$
d\rho \ra \ ^g(d\rho )=dg\rho g^* +dgdg^* +g d\rho g^* -g\rho dg^*
\eqno(2.18)$$
Working in the representation $\pi $, we see from eq.(2.15) that
$$
\pi (\ ^g\rho )=g\pi (\rho )g^* +g[D,g^*] \eqno(2.19)
$$
and this can be written in the form
$$
\pi (\ ^g\rho )=\sum_i \ ^ga^i[D,\ ^gb^i] \eqno(2.20)
$$
As expected, the  Dirac operator is not acted up on by the
gauge transformations (i.e. $\ ^gD\equiv D $). The curvature
$\theta $, defined by
$$
\theta =d\rho +\rho^2  \eqno(2.21)
$$
is easily seen to be covariant under the gauge transformations
$$
\theta \ra \ ^g\theta =g\theta g^*  \eqno(2.22)
$$
To see how gauge transformations act on the components of $\pi (\rho
)$, we first give the representation of $g$:
$$
g\ra {\rm diag }(g_1,g_2,g_3)  \eqno(2.23)
$$
where $g_1$, $g_2$ and $g_3$ are $n\times n$, $p\times p$ and
$q\times q $ unitary matrix-valued functions respectively.
A simple computation, using
the commutator $[D,g]$ in eq.(2.8), gives the component form of eq.(2.19):
$$\eqalign{
\ ^gA_m&=g_mA_mg_m^* +g_m\di g_m^* ,\qquad m=1,2,3 \cr
\ ^g(\phi_{mn}+M_{mn})&=g_m(\phi_{mn} +M_{mn} )g_n^*, \qquad m\not=n
\cr}\eqno(2.24)
$$
In this form it becomes manifest that the $A_m$ are the usual
gauge  fields, while the combinations $\phi_{mn} +M_{mn} $ are
scalar fields transforming covariantly under the mixed gauge
transformations $g_m$ and $g_n$.(We use that  $\ ^g M_{mn}
=M_{mn}$ in $D$ .) The fields $\phi_{mn} $ are the physical fields, and the
$M_{mn}$ are the vacuum expectation values of the Higgs fields.
In other words, the Higgs potential will turn out to have
its minimun when $\phi_{mn}=0$,
indicating that the scalar fields appearing are the fluctuations
around the vacuum state, and that we are in the spontaneously broken
phase. To pass to the symmetric phase, we must reexpress all the
scalar fields in the combination $\phi_{mn}+M_{mn} $.

It was noted by Connes and Lott [4] that the representation $\pi $ is
ambiguous, a fact that will explain the appearence of auxiliary
fields. This can be seen from the fact that if $\pi (\rho )$ is set
to zero $\pi (d\rho ) $ is not necessarily zero, and the correct
space of forms to work on is ${\Omega^* ({\cal A})\over Ker \pi +
d Ker \pi }$, where $Ker \pi $ is the kernel of the map $\pi $.
 Thus the auxiliary fields can be either quotiented out or eliminated
through their equations of motion as they are non-dynamical. We
choose to keep the auxiliary fields explicitly
in our calculations (rather than modding them out)
since  the step of identifying which fields are
genuinely independent is complicated and model-dependent.
However, Proposition 4 in [4] shows that, for the Yang-Mills functional,
the two procedures are equivalent.
Next we proceede to compute $\pi (d\rho )$ which is a lengthy
calculation. The elements of this matrix are functions of the $a^{i'}s$
and the $b^{i'}s$ and must be reexpressed in terms of the fields
$A_m$, $\phi_{mn}$ and possibly new independent fields. We
first consider
$$\eqalignno{
\pi (d\rho )_{11}&=\sum_i \di a_1^i \di b_1^i +\sum_i (M_{12}a_2^i-a_1^iM_{12})
(M_{21}b_1^i-b_2^iM_{21}) \cr
&\qquad +\sum_i (M_{13}a_3^i-a_1^iM_{13})(M_{31}b_1^i-b_3^iM_{31})\cr
&=\di A_1 +M_{12}\phi_{21}+\phi_{12}M_{21}+M_{13}\phi_{31}+\phi_{13}
M_{31}-X_{11} &(2.25)\cr}
$$
where the auxiliary field $X_{11}$ is given by
$$
X_{11}=\sum_i a_1^i \bigl( \di^2 b_1^i +[M_{12}M_{21}+M_{13}M_{31},
 b_1^i]\bigr) . \eqno(2.26)
$$
Before  continuing our calculations, we would like to point out the
 following problem
and   the necessary modifications needed to remedy it. The part
$$\sum_i a_1^i \di^2 b_1^i $$ of the auxiliary field $X_{11}$ is
an $n\times n$ matrix whose elements are arbitrary functions. Thus
the terms of the scalar Higgs potential  could be absorbed in it.
 This, of course,
would be undesirable for any model (since all the scalar fields
would remain massless at the classical level). What saves the potential
from disappearing alltogether is to include the information about the
mixing between the three generations of quarks and leptons
in the Dirac operator. This mixing is related to the fermionic mass
matrix. Therefore the Dirac operator used in eq.(2.3)
should be modified to
$$
D=\pmatrix { \di \bt I\bt I &\g5 \bt M_{12}\bt K_{12} & \g5 \bt M_{13}
\bt K_{13} \cr
\g5 \bt M_{21}\bt K_{21} & \di \bt I \bt I & \g5 \bt M_{23} \bt K_{23} \cr
\g5 \bt M_{31} \bt K_{31} & \g5 \bt M_{32} \bt K_{32} & \di \bt I\bt I \cr
}\eqno(2.27)
$$
where $K_{mn}=K_{nm}^* $.
The matrix $K$ commutes with the $a^i$ and $b^i$. This modification
implies that $\pi (\rho ) $ is obtained by substituting
$$
\phi_{mn} \rightarrow \phi_{mn} \bt K_{mn}  \eqno(2.28)
$$
and $\pi (d\rho )_{11} $ given in eq.(2.25), now becomes \footnote*
{We omit the tensor product signs to simplify notation. Thus, e.g.
$K_{mn} $ means $1\bt 1\bt K_{mn} $ and $M_{mn}$ means
$1\bt M_{mn}\bt 1$.}
$$
\pi (d\rho )_{11}=\di A_1 +\bigl( \vert K_{12} \vert^2(M_{12}\phi_{21}
 +\phi_{12} M_{21})
+\vert K_{13}\vert^2 (M_{13}\phi_{31} +\phi_{13}M_{31} )\bigr)-X_{11}
\eqno(2.28)
$$
where the new field $X_{11} $ is given by
$$
X_{11}=\sum_i a_1^i \bigl( \di^2 b_1^i +[\vert K_{12}\vert^2 M_{12}M_{21}+
\vert K_{13} \vert^2 M_{13}M_{31}, \ b_1^i]\bigr)
\eqno(2.29)
$$
where $\vert K_{ij}\vert^2=K_{ij}^*K_{ij}.$
The other elements of $\pi (d\rho )$ can be found easily and expressed
in the compact and generalizable form
$$
\pi (d\rho )_{mm} =\di A_m + \sum_{n\not =m} \vert K_{mn}\vert^2
(M_{mn}\phi_{nm}
+\phi_{mn}M_{nm} )-X_{mm} \eqno(2.30)
$$
where the $X_{mm}$ fields are defined by
$$
X_{mm}=\sum_i a_m^i \bigl( \di^2 b_m^i +[\sum_{n\not =m}\vert K_{mn}\vert^2
M_{mn}M_{nm}, \ b_m^i]\bigr) .
\eqno(2.31)
$$
The non-diagonal element  $\pi (d\rho )_{12} $ is given by
$$\eqalignno{
\pi (d\rho )_{12} &=\g5 K_{12}\bigl( -\sum_i \di a_1^i (M_{12}
b_2^i-b_1^iM_{12}) + K_{13}K_{32}\sum_i (M_{12}a_2^i -a_1^i M_{12})\di b_2^i
\bigr) \cr &\qquad +\sum_i (M_{13}a_3^i -a_1^i M_{13})(M_{32}b_2^i
-b_3^i M_{32}). &(2.32)\cr}
$$
and can be rewritten in terms of the fields $A_m$ and $\phi_{mn}
$ and a
new field $X_{12}$
$$
\pi (d\rho )_{12} =-\g5 K_{12}\bigl( \di \phi_{12} +A_1 M_{12}-M_{12}
A_2 \bigr) +K_{13}K_{32} \bigl( M_{13}\phi_{32}+\phi_{13}M_{32} - X_{12}\bigr)
\eqno(2.33)
$$
where the new field $X_{12}$ is given by
$$
X_{12}= \sum_i a_1^i (M_{13}M_{32}b_2^i -b_1^i M_{13}M_{32})
\eqno(2.34)
$$
Similarly the other non-diagonal elements may be written in a
compact and generalizable form:
$$\eqalignno{
\pi (d\rho )_{mn}&=-\g5 K_{mn} \bigl( \di \phi_{mn} +A_m M_{mn}
-M_{mn}A_n \bigr) \cr &\qquad + \sum_{p\not= m,n}K_{mp}K_{pn}\bigl(
M_{mp}\phi_{pn}+\phi_{mp}M_{pn}\bigr) - X_{mn}, \qquad m\not= n,
&(2.35) \cr}
$$
where the fields $X_{mn}$ are defined by
$$
X_{mn}= \sum_i a_m^i \sum_{p\not= m,n} K_{mp}K_{pn}\bigl( M_{mp}M_{pn}
b_n^i -b_m^iM_{mp}M_{pn} \bigr) ,\qquad m\not= n, \eqno(2.36)
$$
The elements $\pi (d\rho )_{mn} $ are self adjoint,
$$
\pi (d\rho )_{mn}^{\ast } =\pi (d\rho )_{nm}  \eqno(2.37)$$
Collecting all these results, the representation of the curvature
$\pi (\theta )$ can be written in terms of components. First, the
diagonal elements are given
$$
\pi (\theta )_{mm}={1\over 2}\gamma^{\mu \nu}F_{\mu\nu}^m
+\bigl(\sum_{p\not=m} (\vert K_{mp}\vert^2\vert \phi_{mp} +M_{mp}\vert ^2-
Y_m \bigr)-X'_{mm} \qquad m=1,2,3 \eqno(2.38)
$$
where we have defined
$$\eqalign{
X'_{mm}&=\sum_i a_m^i\di^2 b_m^i +\ (\partial^{\mu}
A_{\mu}^m+A^{\mu m}A_{\mu}^m) \cr
F_{\mu\nu}&=\partial_{\mu}A_{\nu}^m-\partial_{\nu}A_{\mu}^m
+[A_{\mu}^m,A_{\nu}^m]  \cr
Y_m&= \sum_{p\not=m}\sum_i a_m^i \vert K_{mp}\vert^2
\vert M_{mp}\vert^2 b_m^i}\eqno(2.39).
$$

The non-diagonal elements of $\pi (\theta )$ are given by
$(m\not=n )$:
$$\eqalignno{
\pi (\theta )_{mn}&=-\g5 K_{mn}\bigl( \di \phi_{mn}+ A_m (\phi_{mn}+
M_{mn})-(\phi_{mn}+M_{mn})A_n\bigr)-X_{mn} \cr
&\qquad +\sum_{p\not= m,n}  K_{mp}K_{pn}
\bigl( (\phi_{mp}+M_{mp})(\phi_{pn} +M_{pn})
-M_{mp}M_{pn}\bigr) &(2.40)\cr}
$$
where we have used the notation $\vert \phi_{mp} \vert^2=\phi_{mp}\phi_{pm}.$
The curvature is self-adjoint : $\pi (\theta )_{mn}^*=\pi (\theta )_{nm}$.

The fields  $Y_m$ and $X_{mn}$ are not all independent.
Depending on the structure of the mass matrices $M_{mn}$,some of them could be
expressed in terms of $\phi_{mn} $. If it so happens that all
the $X$-fields are independent then after eliminating all the auxiliary
fields, the scalar potential will disappear. This does not happen
if the mass matrices are chosen in such a way  as to correspond to a
possible vacuum with symmetry breaking. In the examples that we consider
here, the potential will survive.

The Yang-Mills action is given by the positive-definite expression
$$
I={1\over 8} Tr_{\omega} \bigl(\theta^2 \vert D \vert^{-4}\bigr)
\eqno(2.41)
$$
where $Tr_{\omega }$ is the Dixmier trace. It is defined by
$$
Tr_{\omega}(\vert T\vert )={\rm lim}_{\omega}{1\over \log N}
\sum_{i=0}^N \mu_i(T)  \eqno(2.42))
$$
where $T$ is a compact opreator, and $\mu_i$ are the eigenvalues
of $\vert T \vert $. This trace effectively picks
out the coefficient of the logarithmic divergences. For the Dirac
operators we shall consider the Dixmier trace to be
 equivalently replaced with
a heat kernel expression, using the identity
$$
\vert D \vert^{-4} =\int_0^{\infty} d\epsilon \epsilon e^{-\epsilon
\vert D \vert^2}. \eqno(2.43)
$$
and the expansion
$$
{\rm tr }\bigl(fe^{-\epsilon \vert D \vert^2} \bigr) = \int d^4x {\sqrt g}f(x)
({a_0\over \epsilon^2}+{a_1\over
\epsilon }+\ldots ) \eqno(2.44)
$$
where $a_0=1$, $g $ is the metric, and $a_1=R$ is the curvature scalar.
This can be used to show
 that the action (2.41) is equal to
$$
I={1\over 8}\int d^4x \sqrt g {\rm Tr}\bigl( {\rm tr} (\pi^2 (\theta )) \bigr)
 \eqno(2.45)
$$
where tr is taken over the Clifford algebra, and Tr is taken over the
matrix structure. Using eqs (2.38) and(2.45) the action takes the
familiar form (in Euclidean space):
$$\eqalignno{
I&=-\sum_{m=1}^3 Tr\Bigl( {1\over 4}  F_{\mu\nu}^m F^{\mu\nu m}
-{1\over 2}\Bigl\vert  \sum_{p\not= m}\bigl(\vert K_{mp}\vert^2
\vert \phi_{mp}+M_{mp}\vert^2
-Y_m \bigr)-X'_{mm}\Bigr\vert^2 \cr
& -{1\over 2} \sum_{p\not=m} \vert K_{mp}\vert^2
\Bigl\vert \partial_{\mu}
(\phi_{mp}+M_{mp})+A_{\mu m}(\phi_{mp}+M_{mp})-(\phi_{mp}+M_{mp})
A_{\mu p}\Bigr\vert^2 \cr
& +{1\over 2} \sum_{n\not=m}\sum_{p\not=m,n}\Bigl\vert
 \vert K_{mp}\vert^2\bigl(
(\phi_{mp}+M_{mp})(\phi_{pn}+M_{pn})-M_{mp}M_{pn}\bigr)-X_{mn}\Bigr\vert^2
\Bigr) &(2.46)\cr}
$$
where we have used the notation $\vert D_{\mu} \phi\vert^2 \equiv
D_{\mu}\phi D_{\nu} \phi g^{\mu\nu}$, and when we analytically continue
to Minkowski space by the change $x_4 \ra it $ the action changes by
$ I_E\ra -I_M $.
This action contains the Yang-Mills action for the gauge fields
$A_{\mu m}$ , kinetic energies for the scalar fields
$\phi_{mn}, m\not=n$, and a potential for the scalar fields.
In the last step, the independent fields from the set $X'_{mm}$, $X_{mn}$, and
$Y_m$ must be eliminated. The result depends on the particular
choices of $M_{mn}$ and is model-dependent.
If the potential survives, it is positive definite
 being a sum of squares and it is minimized for $\phi_{mn}
=0$. Now we are ready to apply this construction to model building.


{\bf\noindent 3. The $SU(2)\times U(1)$ standard model}
\vskip.2truecm

To clarify the general formalism developed in the last section,
we consider the simple example where the  Riemannian
manifold is extended by two points. In all the formulas of the last
section we now set
$$
a_3^i=b_3^i=0, \qquad  M_{13}=M_{23}=0 \eqno(3.1)
$$
We take the elements $a_1 \epsilon M_2({\cal A} )$, and $a_2\epsilon
M_1({\cal A}) $ to be $2\times 2$ and $1\times 1$ matrices, respectively.
The matrix $M_{12} $ is then a $2\times 1$ matrix and will be
chosen to be
$$M_{12}\equiv \mu S =\mu \pmatrix{0\cr 1\cr }  \eqno(3.2)
$$
The choice of $M_{12}$ dictates the breaking mechanism. With
these choices $\pi (\rho )$ takes the form
$$
\pi (\rho )=\pmatrix{ (A_1)_I^J &H^J \cr H_I^* &A_2 \cr} \eqno(3.3)
$$
where $H^I$ is a $2\times 1$ doublet. We shall also impose the
graded tracelessness condition, ${\rm Tr}(\Gamma_1 \pi(\rho ))=0$,
where $\Gamma_1 $ is the grading matrix $\Gamma_1 = {\rm
diag}(1,-1)$. This implies
$$
{\rm Tr }A_1=A_2  \eqno(3.4)
$$
In terms of the elements $a^i$ and $b^i$ the Higgs field $H$ takes the form
$$ H=\mu \sum_i a_1^i(Sb_2^i-b_1^iS) \eqno(3.5)
$$
while the fields $X_{mn}$ and $Y_m$ are given by
$$\eqalign{
X_{12}&=0=X_{21} \cr
Y_1&=\mu^2\sum_i a_1^i T b_1^i \cr
Y_2&=\mu^2 \cr }\eqno(3.6)
$$
where $T$ is the matrix $\pmatrix{0 &0 \cr 0 &1}$. This implies that
the only auxilirary fields are $X'_{11}$, $X'_{22}$, and
 $Y_1$, and these should be
eliminated. This step can be immediately taken and results in
the disappearence of the terms involving these fields. The final action then
takes the form (in Minkowski space ):
$$\eqalignno{
I&=\Bigl( {1\over 4}\bigl( (F_{\mu\nu}^1)_I^J (F_{\mu\nu }^1)_J^I
+(F_{\mu\nu}^2)(F_{\mu\nu}^2) \bigr) \cr
&\qquad +{1\over 2}\k2 \Bigl\vert \partial_{\mu} (H^I+H_0^I) +(A_{\mu 1})_J^I
(H^J+H_0^J)-(H^I+H_0^I )A_{\mu 2} \Bigr\vert^2 \cr
&\qquad -{1\over 2} \l4 \bigl( (H^I+H_0^I)(H_I^*+H_{0I}^* )-\mu^2 \bigr)^2
\Bigr)&(3.7) \cr}
$$
where we have normalized the trace such that ${\rm Tr} 1=1 $.
Note that, for this normalization of the trace, ${\rm Tr}(KK^*)^2-
({\rm Tr }KK^*)^2$ is positive (non-negative), by the Schwarz inequality
for Tr. Thus the coupling constant of the quartic term in the
Higgs potential is non-negative which guarantees stability of the
theory at tree level. We also have that ${\rm Tr }(KK^*)^2-
({\rm Tr }KK^*)^2 \le (n-1)({\rm Tr}KK^*)^2$, where $n$ is the
number of rows and columns of $K$. Therefore, the order of magnitude
of the bare quartic Higgs coupling constant is the same as the order
of magnitude of the square of the bare gauge coupling constant.
The potential is minimised when $H^I=0$. The fields are already expanded
around the vacuum state, and the minimum corresponds to the broken
phase. To display gauge invariance explicitely, the action could
be easily expressed in terms of the shifted field $H+H_0$.

The gauge fields are in the familiar form of the standard model [7],
but in the broken phase. By writing
$$\eqalign{
(A_1)_I^J&= i\pmatrix{A^0+A^3 &A^1-iA^2 \cr
A^1+iA^2 &A^0-A^3\cr }\cr
A_2&=2iA^0 \cr} \eqno(3.8)
$$
one finds that $A_{\mu}^1-iA_{\mu}^2=W_{\mu}$ and
$A_{\mu}^0+A_{\mu}^3=Z_{\mu}$ are the $W$ and $Z$
gauge fields.
The leptons fit naturally in this scheme, and can be included by introducing
the spinors $L$ subject to the chirality condition
$$
\bigl(\g5 \ot \Gamma_1\bigr) L=L \eqno(3.9)
$$
where this condition will only  be imposed after we  have performed
the Wick rotation from Euclidean to Minkowski space. The spinors $L$
then take the form
$$
L=\pmatrix{l_L \cr  e_R^- \cr} \eqno(3.10)
$$
where the left-handed electron and neutrino are in the first copy
and form a doublet of $SU(2)$: $l_L=\pmatrix{\nu_e \cr e^-\cr }_L$,
while the right-handed electron is in the second copy
and is a singlet as can be deduced from the form of the elements
$a^i$ and $b^i$. The  leptonic action is then given by
$$\eqalignno{
I_l&=<L,(D+\pi(\rho ))L> \cr
  &=\int d^4x \ol L\bigl(D+\pi (\rho ) \bigr) L &(3.11a)\cr}
$$
In terms of components this becomes
$$\eqalign{
I_l&=\int d^4x \Bigl[ {\ol l_L} (D +\pi (A))l_L +{\ol e_R} (\di +A) e_R \cr
&\qquad \qquad +{\ol l_L} (H+H_0)e_R K +{\ol e_R} (H^*+H_0^*)
 l_L K^* \Bigr] \cr}\eqno(3.11b)
$$
Thus, as required, the electron becomes massive, while the neutrino
remains massless.

Introducing $SU(3)$ and the quarks is more complicated in this approach.
The reason is that $SU(3)$ is not   broken, and no Higgs
fields are necessary. It can be introduced in an essentially  commutative
way. The solution adopted in [3-4] was to introduce a bimodule.
One must introduce, in addition, a new algebra ${\cal B}$ which must be
taken to be $M_1(C)\bp M_3(C) $. The mass matrices in the Dirac operator
along these directions are taken to be zero, forcing the
vanishing of the Higgs fields along the same directions. Because the
hypercharge
assignments of the quarks are delicate, the different $U(1)$ factors
must be related. This is achieved with the following assignments:
On the algebra ${\cal A}$ we must set ${\rm Tr}A_1 =0 $, $A_2=-Y $
and on the algebra ${\cal B}$ we must set $B_1=-Y=-{\rm Tr}B_2 $.
This prescription  guarantees the correct hypercharge
assignments for the quarks and leptons. The following point is in
order. Although the relation between the different $U(1)$ factors
could be obtained from the mathematical condition of the unimodularity
of the algebras considered, it is clear that this condition is not
natural, especially since the main motivation behind
the non-commutative picture is to explain the geometric origin
of the Higgs fields, and of the phenomena of  symmetry breaking. Introducing
$SU(3)$ in a commutative way and  decoupling it from the rest is
not very convincing. However we shall proceed in our construction
for illustration
and to show that it is perfectly possible to obtain the standard model
using this method.

The quarks are taken to be in the representation
$$
Q=\pmatrix{{1\over \sqrt 2 }q_L\cr d_R \cr } \eqno(3.12)
$$
subject to the chirality condition $\g5 \ot \Gamma_1 (Q)=Q $, and
$q_L=\pmatrix{u_L \cr d_L \cr}$ is a left-handed $SU(2)$ doublet.
Unfortunately, an action similar to that of the leptons will leave the up
quarks
massless. To avoid this one must also introduce the "dual" quark
representation
$$
\tilde Q=\pmatrix{{1\over \sqrt 2}{\tilde q_L} \cr u_R } \eqno(3.13)
$$
where ${\tilde q}=\pmatrix{d_L \cr -u_L \cr }=i\tau_2 q $
is also a doublet of $SU(2)$. By taking the matrix $K$ to be
$$
K={\rm diag }({\rm h}_{\alpha \beta }^e,{\rm h}_{\alpha \beta }^d,
{\rm h}_{\alpha \beta }^u)  \eqno(3.14)
$$
where the  ${\rm h}_{\alpha \beta }$'s are matrices in generation
space. By defining the spinor
$$
\psi^{\alpha }=\pmatrix{L^{\alpha }\cr Q^{\alpha }\cr {\tilde Q}^{\alpha }
\cr} \eqno(3.15)
$$
where $\alpha =1,2,3 $ refer to the three families, the full
fermionic action can then be written as
$$\eqalignno{
I_f&=<\psi ,(D+\pi(\rho ))\psi > \cr
 &=\int d^4x {\ol \psi }(D +\pi (\rho ))\psi &(3.16)\cr}
$$
and when this action is expanded in terms of components, it gives exactly
the fermionic action of the standard model.
This shows that the standard model can
be obtained  within the non-commutative setting.
But as mentioned earlier, the $SU(2)\times
U(1)$ sector fits more naturaly into this formalism than the $SU(3)$
sector.

{\bf\noindent 4. The $SU(5)$ unified theory }
\vskip.2truecm

The way the strong interactions were introduced in the standard model
suggests that a  unified picture is more desirable from the geometrical
point of view. This was also one of the reasons why
model builders constructed unified theories. Another reason is
that it appears to be natural to assume that, at higher energies, the standard
model is replaced by a more unified picture. The simplest
example of such a scheme is the $SU(5)$ gauge theory [8], which is
the lowest rank group containing $SU(3)\times SU(2)\times U(1)$
as a subgroup. The $SU(5)$ theory is spontaneously broken at two scales.
At the grand unification scale $M$, $SU(5)$ is broken to
 $SU(3)\times SU(2) \times U(1)$
and this, in turn, is broken at the weak scale $\mu $ to $U(1)_{em}$.
The role of
scales in non-commutative geometry is  to  measure the distance between
the different copies of the space. Thus to reproduce the $SU(5)$
theory we need to take the space to be the product of the Reimannian
manifold times three points. In the minimal $SU(5)$ theory the first
stage of breaking is achieved through the use of the adjoint Higgs
representation ${\underline 24} $ and the second stage through the
fundamental ${\underline 5}$ representation. The vacuum expectation
value of the adjoint Higgs is taken to be
$$
\Sigma_0 =M {\rm diag }\ (2,2,2,-3,-3)  \eqno(4.1)
$$
which breaks the symmetry from $SU(5)$ to $SU(3)\times
SU(2)\times U(1)$. This is further broken to $U(1)_{em}$ when the
fundamental Higgs acquires the vacuum expectation value
$$
H_0=\mu \pmatrix{0\cr 0\cr 0\cr 0\cr 1\cr }\equiv \mu S \eqno(4.2)
$$

If the Riemannian manifold is extended by three
points, the simplest possibility to obtain two scales and not three,
as well as a Higgs field belonging to the adjoint representaion and
not to a product representation, is to identify two copies. In other
words we must have a permutation symmetry under the exchange
$1\leftrightarrow 2$. Therefore we must identify
$$
a_1^i=a_2^i, \qquad b_1^i=b_2^i   \eqno(4.3)
$$
The operators in the first and second copies must be taken to be
$5\times 5$ matrices. To obtain the fundamental Higgs  fields,
the third copy must correspond to $1\times 1$ matrices, and,
to avoid an extra $U(1)$ factor we take these elements to be real.
Therefore we must consider the algebra
$$
{\cal A}=M_5(C)\bp M_5(C) \bp M_1(R) \eqno(4.4)
$$
With these choices the vector potential $\pi (\rho )$ becomes
$$
\pi (\rho )=\pmatrix{A & \Sigma &H \cr \Sigma &A &H \cr
H^* &H^* &0 \cr}  \eqno(4.5)
$$
where $A_1=A_2=A=\gamma^{\mu}(A_{\mu})_I^J $ is a self-adjoint $5
\times 5$ gauge vector, $\Sigma_I^J $ is a self-adjoint $5\times
5$ scalar field, and $H^I $ is a complex scalar field. The reason
that $\Sigma $ is self-adjoint lies in  the permutation symmetry,
as $\Sigma_{21} =\Sigma_{12} =\Sigma_{12}^* $. This is also the reason
why the $H's$ in the first and second row are equal. The vector $A_3$
vanishes, because the self-adjointness condition implies that
$A_{\mu 3}=-A_{\mu
3}^* $, but as $A_{\mu 3}=\sum_i a_3^i\partial_{\mu} b_3^i $ is real,
it must vanish. The tracelessness condition
$$
{\rm Tr }(\Gamma_1 \pi (\rho ) )=0  \eqno(4.6)
$$
where $\Gamma_1 ={\rm diag }(1,1,-1) $ implies
$$
{\rm Tr } A=0  \eqno(4.7)
$$
reducing $U(5)$ to $SU(5)$.  In our method, the  symmetry
breaking pattern is specified by choosing the mass matrices
$M_{mn}$ in the Dirac operator to correspond to the desired vacuum.
We shall then take
$$\eqalign{
M_{12}&=M_{21}=\Sigma_0 \cr
M_{13}&=M_{23}=H_0 \cr }\eqno(4.8)
$$
Writing the fields explicitly we find (see eq. (2.27)):
$$
\eqalign{
A_J^I &=\sum_i (a_1^i\di b_1^i)_J^I ,\qquad  \sum_i a_1^
ib_1^i=1 \cr
\phi_{12} &\equiv \Sigma_J^I =\sum_i \bigl( a_1^i [\Sigma_0 ,b_1^i]
\bigr)_J^I \cr
\phi_{13} &\equiv H^I =\mu \sum_i \bigl( a_1^i (Sb_3^i-b_1^i S)\bigr)^I
\cr }\eqno(4.9)
$$
To determine the potential by eliminating the auxiliary fields,
we must determine whether the new functions $X_{mn}, m\not=n $,
and $ Y_m $ are independent. First we find
$$
X_{12}={\mu}^2 \sum_i a_1^i [SS^*, \ b_1^i]  \eqno(4.10)
$$
which is equal to $X_{21}$. Clearly this is a new field which cannot
be expressed in terms of the $\Sigma $ and the $H$ and thus
must be eliminated. Next we calculate
$$\eqalignno{
X_{13}&= \sum_i a_1^i (M_{13}M_{32}b_3^i-b_1^iM_{13}M_{32}) \cr
&=-3M H^I &(4.11)\cr}
$$
which is equal to $X_{23}$, and where  we have used the property
that  $\Sigma_0 H_0=-3M H_0 $. Obviously these fields are not auxiliary
and contribute to the potential. The other fields $X_{31},
X_{32} $, are the conjugate of $X_{13}$. The $Y's$ are more subtle.
First, we have that
$$
Y_1=\sum_i a_1^i \bigl( \vert K_{12}\vert^2 \Sigma_0^2
+\vert K_{13}\vert^2 H_0 H_0^* \bigr) b_1^i \eqno(4.12)
$$
which appears to be independent. However, because of the property
of $\Sigma_0 $
$$
\Sigma_0^2={1\over 5} {\rm Tr} \Sigma_0^2 -M \Sigma_0  \eqno(4.13)
$$
and eq.(4.10), and the definition of $\Sigma $:
$\Sigma =\sum_i a_1^i[\Sigma_0 ,\ b_1^i] $,
 we can rewrite  it as
$$
Y_1=\vert K_{12} \vert^2 (-M\Sigma +\Sigma_0^2)
+ \vert K_{13}\vert^2 (H_0 H_0^* +X_{12}) \eqno(4.14)
$$
It thus must be kept, as it will contribute when $X_{12}$ is eliminated.
 Finally $Y_2=Y_1$ and $Y_3= 2\mu^2\vert K_{31}\vert^2 $.
 Collecting all these results,
the action takes the form
$$\eqalignno{
I&={\rm Tr} \Bigl( {1\over 2}F_{\mu \nu}F^{\mu\nu} +
\vert K_{12}\vert^2 \Bigl\vert \bigl(\partial_{\mu} (\Sigma +\Sigma_0 )
 +[A_{\mu},\ \Sigma +\Sigma_0]
\bigr)_J^I\Bigr\vert^2 \cr
&\qquad +\vert K_{13} \Bigl\vert^2
\vert \bigl( (\partial_{\mu} +A_{\mu}) (H+H_0)\bigr)^I
\Bigr\vert^2 -V(H,\Sigma )\Bigr) &(4.15)\cr }
$$
where the potential $V(H, \Sigma )$ is the scalar potential and
is given by
$$\eqalignno{
V&={\rm Tr}\Bigl( \Bigl\vert \bigl( \vert K_{12}\vert^2
(\Sigma +\Sigma_0 )^2 +
\vert K_{13}\vert^2 (H+H_0 )(H+H_0)^*
-Y_1 \bigr)-X^{'}_{11}\Bigr\vert^2 \cr
&\qquad +\Bigl\vert K_{13}^4 \vert \bigl(\vert H+H_0 \vert^2 -H_0 H_0^*
-X_{12}\bigr)_J^I\Bigr\vert^2 \cr
&\qquad +\vert K_{12}K_{23}\Bigl\vert^2
 \vert \bigl( (\Sigma +\Sigma_0 +3M)(H+H_0)\bigr)^I
\Bigr\vert^2 \cr
&\qquad +2\Bigl\vert  K_{31}\vert^4 \bigl( (H+H_0)^*(H+H_0)-\mu^2 \bigr)
-X^{'}_{33} \Bigr\vert^2\Bigr) &(4.16) \cr}
$$
We now eliminate the auxiliary field $X_{12}$
from the first two terms of the potential. This is then followed
by eliminating $X^{'}_{11}$  and $X^{'}_{33} $ to get the
manifestly gauge invariant potential
$$\eqalignno{
V(\Sigma ,H)&=
\bigl( {\rm Tr} \vert K_{12}\vert^4 -({\rm Tr}\vert K_{12}\vert^2)^2\bigr)
\Bigl\vert \bigl( (\Sigma +\Sigma_0 )^2+M(\Sigma +\Sigma_0 )
-{1\over 5} {\rm Tr}\Sigma_0^2 \bigr)_J^I
\Bigr\vert^2 \cr
&\qquad +{\rm Tr }\vert K_{12}K_{23}\vert^2 \Bigl\vert
\bigl( (\Sigma +\Sigma_0 +3M )(H+H_0)\bigr)^I
\Bigr\vert^2 \cr
&\qquad +2\bigl( {\rm Tr }\vert K_{31}\vert^4 -({\rm Tr }
\vert K_{31}\vert^2 )^2\bigr)  \Bigl\vert (H+H_0)^*(H+H_0)-\mu^2 \Bigr\vert^2
 &(4.17) \cr }
$$
Clearly the potential is positive-definite, and the minimum
occurs when $\Sigma =0=H $. Also, in order not to loose the
$\Sigma $ potential, the matrix $K_{12}$ should be different from the
identity matrix.
So the picture we have  is that of a space-time consisting of
three copies where
two of the copies are identical and seperated by a distance of
order $M^{-1}$. These in turn are seperated from the third copy
by a distance of order $\mu^{-1} $.

At the next stage, we introduce the fermions into the picture.
As is known, the fermions fit neatly as left-handed chiral
spinors, in ${\ol 5}+ 10 $
representations of $SU(5)$ denoted by $ \psi_I +\psi^{IJ} $ [8].
The fermionic action contains, besides the kinetic energies
interacting with the $SU(5)$ gauge fields, the Yukawa couplings
$$
I_y= \int d^4x \bigl(f_{\alpha \beta}
{\ol \psi}_{I\alpha}^c H_J^* \psi_{\beta}^{IJ}
+f'_{\alpha \beta }\epsilon_{IJKLM} {\overline \psi_{\alpha}^c}^{IJ}
\psi_{\beta}^{KL} H^M \ +h.c \bigr)  \eqno(4.18)
$$
where $f_{\alpha \beta } $ and $f'_{\alpha \beta }$ are matrices
in the family space, $\alpha ,\beta =1,2,3 $, and $\psi^c =C{\ol
\psi }$ , is the charge conjugate spinor  having the same chirality
as $\psi $, ($C $ being the charge
conjugation matrix).

In the present formulation  of non-commutative geometry, the
full fermionic action including both the kinetic terms and the
Yukawa couplings must be obtained from an expression of the
form $<\Psi ,(D+\pi(\rho ))\Psi >$, where $\Psi $ is some appropriate
representation for the spinors. We wish to incorporate the $({\ol 5}+10)_L $
into one spinor
where we shall use the equivalence ${\ol 5_L}=5_R $.
We define the spinor $\Psi $ which transforms as
$$
\Psi \ra \ ^g\Psi =g\ot g \Psi  \eqno(4.19)
$$
under the antisymmetric tensor product representation of
the group $U({\cal A})$ of unitary elements of the algebra
${\cal A}$ (acting on $h\wedge h )$.
A useful representation for this spinor is $\Psi_{AB}$, where
the indices $A$ and $B$
take the  values $A=I_1,I_2,1$ along the directions of the three spaces.
It must then satisfy

$$
\Psi_{AB}=-\Psi_{BA}.    \eqno(4.20)
$$
This together   with the permutation
symmetry $1\leftrightarrow 2$ implies that
 $\Psi_{AB}$ has the folowing components:
$$\eqalign{
\Psi_{I_1J_1}&=\Psi_{I_2J_2}={1\over \sqrt 6}\psi_{IJ}\cr
\Psi_{I_1J_2}&=\Psi_{I_2J_1}=0 \cr
\Psi_{I_1 1}&=-\Psi_{1 I_1}=\Psi_{I_2 1}={1\over \sqrt 2}\psi_I
\cr}\eqno(4.21)
$$
By further imposing the chirality condition $\bigl(\g5 \Gamma_1\ot \Gamma_1
\bigr)\Psi =\Psi $, which can be written in the form
$$
\g5 (\Gamma_1)_A^{A'}(\Gamma_1)_B^{B'}\Psi_{A'B'}=\Psi_{AB}
\eqno(4.22)
$$
one finds that $\psi_{IJ}$ is left-handed and $\psi_I$ is right-handed:
$$
\psi_{IJ}=\psi_{IJ\ (L)} \qquad \psi_I=\psi_{I\ (R)} \eqno(4.23)
$$

To put it differently,
the fermions fit neatly in one spinor in a representation transforming
under the antisymmetric product
of $U({\cal A}) $. The fermionic
action is then
$$\eqalignno{
I_{1f}&=<\Psi ,(D+\pi (\rho )\ot 1+1\ot \pi (\rho ))\Psi >\cr
&=\int d^4x  {\ol \Psi_{AB}} \bigl( D\Psi_{AB} +\pi (\rho )_A^C \Psi_{CB}
+\pi (\rho )_B^C \Psi_{AC} \bigr)\cr
&=\int d^4x  \bigl({1\over 3} {\ol \psi_{IJ\ (L)}}(\di \psi_{IJ\  (L)} +A_I^K
\psi_{KJ \ (L)}+A_J^K \psi_{IK\  (L)})+{\ol \psi_{I\ (R)}}(\di +A)_I^J
\psi_{J\ (R)} \cr
&\qquad +{1\over \sqrt 3}(K_{13} {\ol \psi_{I\ (R)}}
(H+H_0)^J\psi_{IJ\ (L)} +h.c)\bigr)
&(4.24) \cr}
$$
and where the particle assignments are taken to be
$$
\psi_{IJ\ (L)}=\pmatrix{ 0& u_3^c &-u_2^c & u_1 &d_1 \cr
-u_3^c &0 &u_1^c &u_2 &d_2 \cr
u_2^c &-u_1^c &0 &u_3 &d_3 \cr
-u_1 &-u_2 &-u_3 &0 &e^+\cr
-d_1&-d_2&-d_3&-e^+&0\cr}_L  \eqno(4.25)
$$
for the $10$ representation, and
$$
\psi_I^c=\pmatrix{d_1\cr d_2\cr d_3\cr e^c \cr \nu^c \cr}_R \eqno(4.26)
$$
for the $5$ representation.
It is clear that this interaction provides masses to the leptons
and down quarks
but not to the up quarks. (The neutrinos will be always massless
since they have no right-handed partners). This situation is identical
to the one we faced for the Yukawa couplings of the quarks in the
standard model. In terms of the $SU(5)$ couplings such
Yukawa couplings come
from the second  term in eq(4.18) and require the introduction of the
epsilon  tensor of $SU(5)$.  We then introduce the spinor in the
completely antisymmetric tensor product $
U({\cal A}) $, (acting on $h\wedge h\wedge h $). It can
be represented by a spinor
$\chi^{ABC} $ completely antisymmetric in $A, B, C$.
 Because of the permutation
symmetry $1\leftrightarrow 2$ the non-vanishing components are
$\chi^{I_1J_1K_1}=\chi^{I_2J_2K_2}\equiv \chi^{IJK} $ and
$\chi^{I_1J_11}=\chi^{I_2J_21}\equiv\chi^{IJ} $.The chirality condition
on $\chi $ is
$$
\bigl( \g5 \Gamma_1 \ot \Gamma_1 \ot \Gamma_1 \bigr)\chi =\chi .\eqno(4.27)
$$
This implies that
$$
\chi^{IJK}=\chi_{(L)}^{IJK} \qquad \chi^{IJ}=\chi_{(R)}^{IJ}
\eqno(4.28)
$$
 However, since  we do not wish to introduce more particles,
 these spinors will be
related to $\Psi_{AB} $ by the identification
$$\eqalignno{
\chi^{IJK}_{(L)}&={1\over 6\sqrt 2}\epsilon^{IJKMN}\psi_{MN \ (L)}\cr
 \chi^{IJ}_{(R)}&={1\over \sqrt 6} C{\ol \psi_{IJ\ (L)}}&(4.29)\cr}
$$
The action corresponding to the $\chi $-spinor is then
$$\eqalignno{
I_{2f}&=<\chi ,(D+\pi (\rho )\ot 1\ot 1+1\ot \pi (\rho )\ot
 1 +1\ot 1\ot \pi (\rho ))\chi>\cr
&=\int d^4x {\ol\chi^{ABC}}\bigl(
D\chi^{ABC}+3\pi (\rho )^A_E \chi^{EBC} \bigr) &(4.30)\cr}
$$
The component form of this action then takes the form
$$\eqalignno{
I_{2f}&= \int d^4x \Bigl( {2\over 3}{\ol \psi_{IJ\ (L)}}(\di \psi_{IJ\ (L)}
+2(A)_I^K \psi_{KJ\ (L)})\cr
 &\qquad +{1\over 6\sqrt 3} \bigl(K_{31}
\epsilon^{IJKMN}\psi_{IJ}^c (H^*_K+H^*_{0K}) \psi_{MN}+h.c \bigr)
\Bigr)&(4.31) \cr}
$$
Notice that we have defined the spinors in such a way that
their kinetic energies are properly normalized. Finally,
in order to give different masses to the up and down quarks, we must
introduce the spinor
$$
\lambda =\pmatrix {\Psi\cr \chi \cr }  \eqno(4.32)
$$
and take the matrix $K_{13}$ to be of the form
$$
K={\rm diag}({\rm f}_{\alpha \beta },{\rm f'}_{\alpha \beta })  \eqno(4.33)
$$
where ${\rm f}_{\alpha \beta }$ and ${\rm f'}_{\alpha \beta }$ are matrices
in generation space.
In this way the fermionic action could be written compactly
in the form
$$
<\lambda ,(D+\pi (\rho ))\lambda > \eqno(4.34)
$$
Note that the fermionic mass matrix is proportional to $K_{13} $,
while $K_{12}$ is necessary for the survival of the $\Sigma $ self
couplings in the potential.

To summarize, the present picture looks very attractive:
The discrete structure of space becomes apparent, first, at the weak
scale, and the Higgs field $H$ is associated with the mediation
between the third copy and the two identical
copies which, at that scale, would appear to coincide. As we climb
up in energy, probing the smaller distance scale, we encounter the
Higgs field $\Sigma $ associated with the mediation between the two identical
copies. The fermions fit into one representation (and its conjugate)
and their action takes a very simple form. Of course, from the phenomenological
point of view, the gauge group $SU(5)$ has a serious drawback
connected with proton decay. The rate predicted by this model is ruled out
experimentally, and only in more complicated models  this
problem is avoided. The analysis of such models,
 however, is beyond the scope of
this paper.  The construction of a phenomenologically successful model
will be left to the future. Here we content ourselves with the
construction of some prototype models, in order to master
and illustrate the new techniques
advanced here.

{\bf \noindent 5. Left-right $SU(2)_L\times SU(2)_R\times U(1)_{B-L}$
 symmetric model }
\vskip.2truecm
Another class of attractive models which are of phenomenological
interest is the left-right symmetric models. The simplest one of
which is the $SU(2)_L\times SU(2)_R \times U(1)_{B-L}$ theory [9].
The non-commutative geometry setting is perfectly appropriate for product
groups. The Higgs fields used in the breaking are usually taken to
be a $(2,2)$ and $(3,1)+(1,3)$ with respect to $SU(2)_L\times SU(2)_R $.
It is also possible to replace the $(3,1)+(1,3)$ by doublets $(2,1)
+(1,2)$. But this choice is less prefered for phenomenological reasons [9].
As we have learned from the $SU(5)$ theory in order to get an adjoint
Higgs representation, two copies must be identified, i.e. interchanged by
permutation symmetry. Thus, for each $SU(2)$, the Riemannian manifold
should be extended by two points. The total extension is
by four points.
 One can immediately see that  the algebra must be taken to be
$$
{\cal A}_2 =M_2(C) \bp M_2(C)\bp M_2(C)\bp M_2(C) \eqno(5.1)
$$
The elements $a_1, b_1, a_2, b_2 ,a_3, b_3, a_4, b_4 $,
are $2\times 2$
matrices. We must require the permutation symmetries
$1\leftrightarrow 2$ and $3\leftrightarrow 4 $. Then
$$\eqalign{
a_1&=a_2 \qquad b_1=b_2 \cr
a_3&=a_4 \qquad b_3=b_4 \cr}\eqno(5.2)
$$
 The vector potential
$\pi (\rho )$ takes the form
$$
\pi (\rho )=\pmatrix{A_1 &\Delta_1  &\Phi
&\Phi  \cr
\Delta_1 & A_1 & \Phi &\Phi \cr
\Phi^* &\Phi^* &A_2 &\Delta_2\cr
\Phi^* &\Phi^* &\Delta_2 &A_2 \cr} \eqno(5.3)
$$
where $A_1$ and $A_2$ are $U(2)_L$ and $U(2)_R$ gauge fields,
$\Delta_1 $ and $\Delta_2 $ are triplets in the adjoint
representaions of the
respective groups, and $\Phi $ is $(2,2)$ with respect to the product groups.

The mass matrices entering the Dirac operator are taken to be
$$\eqalign{
M_{12}&= M_{21}= \pmatrix {0 &0 \cr v_1 &0 \cr}\equiv v_1 S  \cr
M_{13}&=M_{14}=M_{23}=M_{24}=\pmatrix {u_1 &0 \cr 0 &u_2 \cr}\cr
M_{34}&=M_{43}=\pmatrix {0& 0\cr v_2 &0 \cr }\cr}\eqno(5.4)
$$
where $u_1,v_1,v_2$ are taken to be real, and $u_2$ is taken
to be complex, with the phase of $u_2$  related to CP violation.
 To reduce the gauge group from
$U(2)_L\times U(2)_R $ to $SU(2)_L\times SU(2)_R
\times U(1)_{B-L} $ we impose the tracelessness condition
$$
{\rm Tr}(\Gamma_1 \pi (\rho ) )=0  \eqno(5.5)
$$
where $\Gamma_1={\rm diag}(1,1,-1,-1) $.
 The scalar fields in
$\pi (\rho )$ are given in terms of the $a^i $ and $b^i $ by
$$\eqalign{
\Delta_1 &=v_1\sum_i a_1^iSb_1^i \cr
\Delta_2 &=v_2\sum_i a_3^iSb_3^i \cr
\Phi &=\sum_i a_1^i (M_{13}b_3^i -b_1^i M_{13} )\cr}\eqno(5.6)
$$
These expressions are important in determining which of the
auxiliary fields are independent. The $X$- and $Y$- fields are
given by
$$\eqalign{
X_{12}&=2 (\vert u_2 \vert^2 -\vert u_1 \vert^2 )
\sum_i a_1^i [T,\ b_1^i] \cr
X_{13}&=u_1 \sum_i a_1^i(v_1 Sb_3^i -b_1^iv_2^* S^*) \cr
X_{34}&=(\vert u_2\vert^2-\vert u_1 \vert^2 )\sum_i a_3^i
[T,\ b_3^i] \cr
Y_1&=\k2 \Bigl( 2\vert u_1\vert^2 +
(2\vert u_2 \vert^2-2\vert u_1\vert^2 +\vert v_1\vert^2)
\sum_i a_1^iTb_1^i \ \Bigr)\cr
Y_3&=\Bigl( 2\vert u_1 \vert^2 +(2\vert u_2\vert^2 -2\vert u_1\vert^2
+\vert u_1\vert^2 )\sum_i a_3^iTb_3^i\  \Bigr)\cr}\eqno(5.7)
$$
where the other functions could be determined from the above
using the permutation symmetry, and the matrix $T$ is given
by: $T=\pmatrix{0&0\cr 0&1\cr}$.   For simplicity we have assumed
that $K_{12}=K_{34}=K_{13}=K$.

{}From the form of the action in eq.(2.46)  it is clear that the
field $X_{13}$ is auxiliary and eliminating it will remove the
whole term that appears in $\theta_{13} $. The remaining potential
is given by
$$
V= V_1(\Phi ,\Delta_1) +V_2(\Phi ,\Delta_2 ) \eqno(5.8)
$$
where the two parts are
$$
\eqalignno{
V_1(\Phi ,\Delta_1)&=
2\l4\Bigl\vert 2\vert \Phi +M_{13} \vert^2 +\vert \Delta_1 +M_{12}
\vert^2 \cr
&\qquad -2\alpha Z_1 -\vert M_{12}\vert^2 -2\vert M_{13}\vert^2 \Bigr\vert^2
\cr &\qquad +8{\rm Tr}(KK^*)^2\vert \Bigl\vert
 \Phi +M_{13}\vert^2-\vert M_{13}\vert^2
-\beta Z_1 \Bigr\vert^2  &(5.9a)\cr}
$$
where $\alpha =2\vert u_2 \vert^2 -2\vert u_1\vert^2+\vert
v_1\vert^2 $ and $\beta =\vert u_2\vert^2-\vert u_1\vert^2 $
and $Z_1=\sum_i a_1^i Tb_1^i $ is an auxiliary field. The second
part $V_2$ has a similar structure  and can be obtained
by the substitutions
$$
V_2(\Phi ,\Delta_2 )=V_1 (\Phi \ra \Phi^* ,\Delta_1 \ra \Delta_2 ,
v_1\ra v_2 ,Z_1\ra Z_2) \eqno(5.9b)
$$
where $Z_2=\sum_i a_3^iTb_3^i $. Elimination of $Z_1$ and $Z_2$ will
yield a potential of the desired form.

The leptons have the form
$$
\Psi =\pmatrix{\psi_1\cr \psi_1 \cr\psi_2 \cr \psi_2 \cr} \eqno(5.10)
$$
where  $\psi_1$ and $\psi_2$ are doublets under the two
$SU(2)$ groups.
After imposing the chirality condition
$$
(\g5 \ot \Gamma_1 )\Psi =\Psi \eqno(5.11)
$$
one gets
$$
\psi_1=\psi_{1\ (L)} \qquad  \psi_2=\psi_{2\ (R)} \eqno(5.12)
$$
By writing $\psi=\pmatrix{\nu_e\cr  e^-\cr}$ one finds that
the usual leptons (with  neutrinos acquiring Majorana masses )
emerge. The required coupling is
$$
<\Psi ,(D+\pi (\rho ))\Psi >=\int d^4 x {\ol \Psi }(D+\pi (\rho )
)\Psi \eqno(5.13)
$$
However, in order to make the right fermions heavy, one must introduce
the conjugate fermions
$$
\chi =\pmatrix{\chi_1 \cr \chi_1 \cr \chi_2 \cr \chi_2 \cr}
\eqno(5.14)
$$
required to also satisfy the chirality condition. We shall  make
the identifications,
 $\chi_1=i\tau_2 \psi_1 $ and $\chi_2=i\tau_2 \psi_2 $. The other
term needed in the action is
$$
<\chi ,(D+\pi(\rho ))\chi >=\int d^4 x {\ol \chi }(D+\pi (\rho ))\chi
\eqno(5.15)
$$
and provides, in addition to the kinetic terms which appear again,
the coupling of the conjugate Higgs.  The quarks
can be introduced in a similar manner. However, the correct coupling
to  $U(1)$ can only be achieved after introducing $SU(3)$. This
can be done in a way identical to that in the standard model, where
the $U(3)\times U(1)$ group is coupled through the bimodule structure
with both $U(1)$ and ${\rm Tr} U(3)$  related to the $U(1)_{B-L}$
in order to provide the correct hypercharge assignments for the quarks.
The phenomenological details of this model will be treated elsewhere.

{\bf \noindent 6.Summary and conclusion }
\vskip.2truecm

We have achieved the main objective set for this paper: The construction
of a formalism using the framework of non-commutative geometry
of Alain Connes [1-2]. We have modified one particular point in that
we choose the basic algebra to be a direct sum of matrix algebras.
This simplifies the computations and makes it possible
to consider large groups, while in the original setting
this becomes a very complicated task, since all the elements in the curvature
have to be
computed one by one. Another improvement is
choosing the vacuum state of the potential to appear in the Dirac
operator. Such a choice might appear to break gauge invariance. But
this is not so, as the Dirac operator does not transform under gauge
transformations, and the actions constructed are shown to be
gauge invariant. This is also seen in detail in the component form
of the actions. We have derived the formulas for the Yang-Mills
action corresponding to a continuous space-time multiplied by three points,
but have written the formulas in a way applicable to a space extended
by $N$ points. We have studied three examples in great detail,
and showed, in a step by step calculation, how to extract the bosonic
action by eliminating the auxiliary fields, and how to introduce
fermions in a realistic way. Of course, this formalism does not pretend
to solve the fundamental problem of explaining the fermionic
mass matrices, and thus does not reduce the number of parameters
associated with the fermion masses. It however specifies the Higgs
sector and reduces the number of possible terms at the tree level,
since the potential takes a very specific form. At the classical
level, the cosmological constant always  naturally comes out to be
zero. All  this provides a very good motivation to investigate
some of the problems arising in this formalism.
The first question that one may ask is on how restrictive this
new formalism is. Obviously it is somewhat restrictive,
but not to the point that only few models remain.
One common feature is that the
models that one can construct favor the minimal Higgs representations.
Indeed, if in the $SU(5)$ example one wanted to introduce the Higgs
representation ${\underline {45}}$ one finds that this can only be
done by taking it as an external scalar field not associated
to any vector. Of course, this would be self-defeating and cannot be
considered to be natural.  In this respect, an $SO(10)$ model which
is acceptable phenomenologically is not easy to construct, since
it would require complicated Higgs's such as ${\underline {120} }$ or
${\underline {16}_s}$.

The second question that require further study is the question of
whether space-time supersymmetry can be embedded in non-commutative
geometry. This question does not appear to have an
obvious answer. The difficulty
is that the basic building block in non-commutative geometry is the
Dirac operator, while in supersymmetry, what appears to be more fundamental
is the supersymmetric covariant derivative, $D_{\alpha } $,which
satisfies
$$
\{ D_{\alpha } ,D^{\beta }\}=(\di )_{\alpha}^{\beta }
$$
It is like a square root of the Dirac operator. It would be
extremely interesting if the second question could be answered
in the affirmative
and will have positive consequences for constructing models
which are acceptable phenomenologically.

Most fundamental, however, is the question of quantization of theories
in non-commutative geometry. At present we only have information
about the classical action, and any quantum effects can only be
dealt with by starting from the action extracted in the classical
limit. The non-commutative geometry setting advocated in this paper
(if preserved after quantization ) imposes certain constraints on
the counterterms admissible in the renormalization of the quantum
theories. It is likely that these constraints yield relations between
the square of gauge coupling constants and certain quartic Higgs
coupling constants.
 We hope to report on some of these questions in future projects.

\vskip1truecm
{\bf\noindent Acknowledgments}\hfill\break
We would like to thank D. Wyler for very useful discussions.
\vfill
\eject

{\bf \noindent References}
\vskip.2truecm
\item{[1]} A. Connes, {\sl Publ. Math. IHES} {\bf 62} 44 (1983).

\item{[2]} A. Connes, in {\sl the interface of mathematics
and particle physics }, Clarendon press, Oxford 1990, Eds
D. Quillen, G. Segal and  S. Tsou

\item{[3]} A. Connes and J. Lott,{\sl Nucl.Phys.B Proc.Supp.}
{\bf 18B} 29 (1990), North-Holland, Amsterdam.

\item{[4]} A. Connes and J. Lott, {\sl to appear in Proceedings of
1991 Summer Cargese conference}.

\item{[5]} D. Kastler , Marseille preprints

\item{[6]} R. Coquereaux, G. Esposito-Far\'ese, G. Vaillant,
{\sl Nucl. Phys.}{\bf B353} 689 (1991);\br
M. Dubois-Violette, R. Kerner, J. Madore, {\sl J. Math.
Phys.}{\bf 31} (1990) 316;\br
B. Balakrishna, F. G\"ursey and K. C. Wali, {\sl Phys. Lett.}
{\bf 254B} (1991) 430.

\item{[7]} S. Glashow {\sl Nucl. Phys.}{\bf 22},579 (1961);\br
A. Salam and J. Ward {\sl Phys. Lett}{\bf 13},168 (1964);\br
S. Weinberg, {\sl Phys. Rev. Lett.}{\bf 19}, 1264 (1967);\br
A. Salam, in {\sl Elementary Particle Theory} (editor N. Svartholm),
Almquist and Forlag, Stockholm.

\item{[8]} H. Georgi and S. Glashow, {\sl Phys. Rev. Lett}{\bf 32}
438 (1976);\br
For a review of unified theories see the book by G. Ross
{\sl Grand Unified Theories}, Frontiers in Physics Series,
vol {\bf 60}, Benjamin Publishing.

\item{[9]} R. Mohapatra and J. Pati, {\sl Phys. Rev.}{\bf D11}
566 (1975);\br
 R. Mohapatra and G. Senjanovich {\sl Phys. Rev.}{\bf
D21}165 (1981);\br
 For a review of left-right symmettric models see
the book by R. Mohapatra {\sl Unification and Supersymmetry}
Springer-Verlag, Berlin.

\end